\documentclass[12pt]{iopart}
\usepackage{graphicx}
\usepackage{url}
\begin{document}

\title[The Parkes Pulsar Timing Array]{The Parkes Pulsar Timing Array}

\author{G. Hobbs}

\address{CSIRO Astronomy and Space Science,
PO Box 76, Epping, NSW 1710,
Australia}
\ead{george.hobbs@csiro.au}
\begin{abstract}
The aims of the Parkes Pulsar Timing Array (PPTA) project are to 1) make a direct detection of gravitational waves, 2)  improve the solar system planetary ephemeris and 3) develop a pulsar-based time scale.  In this article we describe the project, explain how the data are collected and processed and describe current research.  Our current data sets are able to place an upper bound on the gravitational wave background that is the most stringent to date. 
\end{abstract}


\section{Introduction}

Approximately two thirds of the available observing time with the Parkes radio telescope is currently dedicated to searching for and studying pulsars.  Numerous pulsars are observed over many years enabling studies of the pulsars themselves, theories of gravity, the interstellar medium and many other phenomena. 

Traditionally pulsars were analysed individually.  The pulsar timing method, in which the pulse arrival times are compared with predictions for those arrival times, is used to obtain accurate measurements of each pulsar's pulse, astrometric and binary parameters.  The resulting post-fit timing residuals indicate unmodelled physical effects that affect the pulsar signal.   Some of these, such as intrinsic instabilities in the pulsar rotation will be specific to a given pulsar.  If the post-fit timing residuals for all pulsars are identical then the cause must be an Earth-based phenomenon.  Processing the data sets with an imprecise knowledge of the solar system planetary masses will affect some pulsars, but not others (depending for instance, on the ecliptic latitude of the pulsar).  A supermassive black hole binary system emitting gravitational waves will induce timing residuals dependent upon the angle between the pulsar, Earth and gravitational-wave source.  Therefore by identifying the angular distribution of the correlations it is possible to disentangle many of these phenomena. This leads to the concept of a Pulsar Timing Array (see, e.g., Foster \& Backer 1990) in which a large number of millisecond pulsars are observed, their post-fit timing residuals determined and a search is made for correlated timing residuals.

Since the year 2005, the Parkes Pulsar Timing Array (PPTA) project  team have been observing a sample of pulsars in order to form a PTA data set. The major scientific aims of the project are to:

\begin{enumerate}
\item make a direct detection of ultra-low-frequency gravitational waves,
\item improve the Solar System planetary ephemeris,
\item develop a pulsar-based time scale.
\end{enumerate}

As described in this review article, significant progress has been made towards all these goals.   A recent detailed summary of the PPTA project has been published in Manchester et al. (2013). Here we provide a brief review of the project as a whole, before describing recent developments that were not described in the earlier publication.

In \S\ref{sec:history} we provide a brief history of the PPTA project. In \S\ref{sec:telescope} we describe the observations.  \S\ref{sec:toad} explains how we process the raw data sets to form pulse times-of-arrival and from them obtain timing residuals.  Our recent science results, current research and future plans are reviewed in the final three sections.

\section{History of the Parkes Pulsar Timing Array}\label{sec:history}


A history of the PPTA project has been presented in Hobbs et al. (2012a). Here we provide a brief summary.  

The first request for observations with the Parkes telescope for the PPTA was submitted in late 2003 and the first high-quality observations were obtained in March 2005.   At that time the basic concept of a PTA was understood, but the necessary timing precision and the total data span required to produce scientifically valuable data was only roughly known.  The initial sample of pulsars was based on the analysis of Jenet et al. (2005) who demonstrated that an isotropic, stochastic gravitational wave background with the amplitude predicted from the available theoretical calculations could be detected if $\sim 20$ pulsars were timed weekly over a period of five years with an rms timing residual of $\sim$100\,ns.  As data were collected, a few pulsars were dropped from the sample and, as new discoveries made, new pulsars added.

Jenet et al. (2006) published the first major result from the PPTA.  That work provided, at the time, the most stringent upper bound on the existence of a gravitational wave background, but only made use of a small subset of pulsars.  The algorithm developed in 2006 could only be applied to pulsar timing residuals that were ``white". However, ``red" (low-frequency) noise was already detectable in many of the data sets.  You et al. (2007a,b) showed that much of this red noise could be attributable to dispersion measure variations from the interstellar medium and/or the solar wind.  This work highlighted the necessity for an observing system that provides sufficient frequency coverage to enable correction for these dispersion measure variations. 

The first analysis of the sensitivity of a PTA to individual, continuous sources of gravitational waves was presented by Yardley et al. (2010) using PPTA data sets. This led to a sky-averaged constraint on the merger rate of nearby ($z < 0.6$) black hole binaries with a chirp mass of $10^{10}$\,M\,$_\odot$ of less than one merger every seven years.  In the year 2010, progress was also made towards the second scientific aim of the PPTA project.  Champion et al. (2010) developed algorithms that allowed errors in the masses of known solar system planets to be identified and this work led to the most precise published estimate for the mass of the Jovian system of $9.547921(2) \times 10^{-4}$\,M\,$_\odot$.

\begin{table}
\caption{Parameters of the current observing systems for the Parkes Pulsar Timing Array}\label{tb:parkes}
\begin{center}
\begin{tabular}{ll}
\hline
Parameter & Value \\ \hline
Telescope diameter & 64\,m \\
10\,cm observing band & 2588 -- 3612\,MHz\\
20\,cm observing band & 1241 -- 1497\,MHz\\
50\,cm observing band & 700 -- 764\,MHz \\
10\,cm system equivalent flux density & 50\,Jy \\
20\,cm system equivalent flux density & 36\,Jy\\
50\,cm system equivalent flux density & 62\,Jy \\
Incoherent digital filter bank systems  & PDFB3, PDFB4 \\
Coherent de-dispersion systems & APSR, CASPSR \\
Typical observing time & 1\,hr\\
\hline
\end{tabular}
\end{center}
\end{table}

Yardley et al. (2011) attempted to make a detection of gravitational waves with the PPTA data sets.  His algorithms were able to detect simulated gravitational waves, but his work showed that our observations were consistent with the hypothesis that no gravitational wave background is present in the data.  However, his algorithm is not effective in the presence of significant red noise in the pulsar timing residuals.  This led to Coles et al. (2011) which described how pulsar data sets should be analysed when affected by red noise.
 
 The third of our major project aims was achieved by Hobbs et al. (2012b) who developed the first pulsar-based time scale that had a precision comparable to terrestrial time standards.   This work demonstrated that, for existing data sets, the atomic timescales were sufficient for our purposes, but with improved data sets it is expected that pulsar-based time scales will become more important. 
 
 Even after the work of You et al. (2007a), the effects of dispersion measure variations were still not fully dealt with.  Keith et al. (2012) developed a new algorithm for the measurement and removal of dispersion measure variations and this is now being routinely applied in our data analysis.
 
 Ravi et al. (2012) improved our prediction of the expected amplitude of a gravitational wave background.  Interestingly, his predictions (based on the Millennium simulation) are ruled out by a new PPTA upper bound on the gravitational wave background (Shannon et al., submitted).  The ramifications of the Shannon et al. limit on the predictions of the expected gravitational wave signal are still being considered.
 
\section{The telescope and details of observations}\label{sec:telescope}


All observations are obtained using the 64-m Parkes radio telescope situated in New South Wales, Australia. Since the year 2005, observations have been made in sessions at 2-3 week intervals.  In each observing session each pulsar is observed at least once in the 20\,cm observing band and once with a dual-frequency 10/50\,cm receiver.   It is often possible within an observing session to obtain more than one observation of each pulsar (particularly for pulsars such as PSR J0437$-$4715 that are situated out of the Galactic plane).  Before each observation a calibration source is recorded to enable subsequent polarisation calibration.  Observations of Hydra A are made at least once during the observing session in order to calibrate the flux density of the pulsar observations.  We currently record the data using four backend instruments: PDFB3, PDFB4, CASPSR and APSR. CASPSR and APSR are coherent dedispersion systems whereas the digital filter bank systems (PDFB3 and PDFB4) do not coherently dedisperse the data.  The current parameters for the observing system are given in Table~\ref{tb:parkes}. Further details about the backend instrumentation are given in Manchester et al. (2013).

The sample of pulsars evolves as new pulsars are discovered. In Table~\ref{tb:pulsars} we list the pulsars that have been observed since Jan 2012 along with their pulse periods, dispersion measures and orbital period.  We also list the number of observations during this time obtained in the 20\,cm and in the 10/50\,cm observing bands (note that 26 observing sessions have been carried out during this time with the first around 5th Jan 2012 and the last 12th April 2013). Many of the pulsars scintillate strongly.  Typical observing durations are 1\,hr, but if the pulsar is weak in one observation we usually stop the observation, move to another pulsar and then, later in the observing session, return to observe the initial pulsar again.  In Table~\ref{tb:pulsars} we list the minimum and median arrival time uncertainties for each observing band.  

\begin{table}
\caption{Parkes Pulsar Timing Array pulsar sample and summary of data sets since Jan. 1 2012.}\label{tb:pulsars}
\begin{footnotesize}
\begin{tabular}{lllllllllllllllllll}
\hline
PSR J & P & DM & ${\rm P}_b$ & \multicolumn{3}{c}{10\,cm} & \multicolumn{3}{c}{20\,cm} & \multicolumn{3}{c}{50\,cm} \\
 &  & &  & N$_{\rm pts}$ & $\sigma_{\rm min}$ & $\sigma_{\rm med.}$ & N$_{\rm pts}$ & $\sigma_{\rm min}$ & $\sigma_{\rm med.}$ & N$_{\rm pts}$ & $\sigma_{\rm min}$ & $\sigma_{\rm med.}$\\
 & (ms)&  (cm$^{-3}$pc)& (days)& & ($\mu$s) & ($\mu$s)& & ($\mu$s) & ($\mu$s)& & ($\mu$s) & ($\mu$s)\\
\hline
J0437$-$4715 & 5.757 & 2.64 & 5.74 & 49 & 0.027 & 0.031 & 208 & 0.028 & 0.038 & 73 & 0.051 & 0.121 \\
J0613$-$0200 & 3.062 & 38.78 & 1.20 & 31 & 1.586 & 2.213 & 47 & 0.397 & 0.729 & 34 & 0.294 & 0.484  \\
J0711$-$6830 & 5.491 & 18.41 & --- & 35 & 0.964 & 3.932 & 53 & 0.503 & 2.072 & 36 & 0.476 & 2.330 \\
J1017$-$7156 & 2.339 & 94.23 & 6.51 & 45 & 0.503 & 2.465 & 60 & 0.122 & 0.331 & 45 & 0.258 & 0.509 \\
J1022$+$1001 & 16.453 & 10.25 & 7.81 & 38 & 0.560 & 1.261 & 48 & 0.121 & 0.789 & 43 & 0.396 & 2.150 \\ \\

J1024$-$0719 & 5.162 & 6.49 & --- & 39 & 1.720 & 3.996 & 47 & 0.216 & 0.163 & 40 & 0.451 & 4.142 \\
J1045$-$4509 & 7.474 & 58.17 & 4.08 & 29 & 4.196 & 6.825 & 36 & 1.072 & 1.629 & 29 & 1.729 & 2.165 \\
J1600$-$3053 & 3.598 & 52.33 & 14.35 & 28 & 0.381 & 0.612 & 40 & 0.209 & 0.273 & 34 & 1.193 & 1.433 \\
J1603$-$7202 & 14.842 & 38.05 & 6.301 & 28 & 1.234 & 8.498 & 40 & 0.447 & 0.892 & 31 & 0.815 & 1.850  \\
J1643$-$1224 & 4.622 & 62.41 & 147.02 & 26 & 1.079 & 1.385 & 36 & 0.469 & 0.597 & 26 & 1.169 & 1.269 \\ \\

J1713$+$0747 & 4.570 & 15.99 & 67.83 & 28 & 0.060 & 0.258 & 41 & 0.021 & 0.110 & 30 & 0.337 & 0.762 \\
J1730$-$2304 & 8.123 & 9.62 & --- & 29 & 0.581 & 2.701 & 31 & 0.320 & 0.895 & 30 & 0.676 & 1.533  \\
J1744$-$1134 & 4.075 & 3.14 & --- & 33 & 0.122 & 0.677 & 40 & 0.073 & 0.319 & 33 & 0.134 & 0.574 \\
J1824$-$2452A & 3.054 & 120.50 & --- & 18 & 0.673 & 1.102 & 28 & 0.168 & 0.255 & 20 & 0.481 & 0.589 \\
J1857$+$0943 & 5.362 & 13.30 & 12.33 & 26 & 0.679 & 2.815 & 34 & 0.368 & 1.037 & 26 & 1.679 & 2.454  \\ \\

J1909$-$3744 & 2.947 & 10.39 & 1.53 & 34 & 0.040 & 0.141 & 59 & 0.010 & 0.072 & 40 & 0.056 & 0.180  \\
J1939$+$2134 & 1.558 & 71.04 & --- & 28 & 0.079 & 0.205 & 37 & 0.015 & 0.035 & 31 & 0.044 & 0.059\\
J2124$-$3358 & 4.931 & 4.60 & --- & 31 & 4.709 & 7.475 & 40 & 0.697 & 1.849 & 33 & 0.519 & 2.979 \\
J2129$-$5721 & 3.726 & 31.85 & 6.63 & 32 & 4.437 & 25.942 & 45 & 0.261 & 1.813 & 33 & 0.321 & 1.204 \\
J2145$-$0750 & 16.052 & 9.00 & 6.84 & 31 & 0.503 & 1.132 & 40 & 0.086 & 0.542 & 35 & 0.345 & 1.125 \\ \\

J2241$-$5236 & 2.187 & 11.41 & 0.15 & 36 & 0.321 & 0.595 & 49 & 0.050 & 0.154 & 38 & 0.033 & 0.220 \\
\hline
\end{tabular}
\end{footnotesize}
\end{table}

\section{Forming pulse arrival times and timing residuals}\label{sec:toad}

The ``raw" data as obtained from the observing system are available for download from the Parkes pulsar data archive (http://data.csiro.au; Hobbs et al. 2011).  An automated pipeline based on the PSRCHIVE software suite (Hotan et al. 2004) runs the following routines on the raw data (details are provided in Manchester et al. 2013):
\begin{itemize}
\item The edges of the observing band are removed along with any identified radio-frequency interference 
\item The best available timing model for the pulsar is installed into the data file
\item Polarisation and flux calibration routines are applied
\item The data are averaged in time, frequency and polarisation to produce a single profile for each observation.\footnote{For PSR~J0437$-$4715 we form and subsequently analyse the invariant interval profile.}
\end{itemize}

The pulse arrival time is then calculated by cross-correlating, in the frequency domain, a noise-free, analytic template with the calibrated observation.

Timing residuals are formed using the \textsc{tempo2} (Hobbs, Edwards \& Manchester 2006) software package. Each arrival time is referred to Terrestrial Time as realised by the Bureau International des Poids et Mesures (BIPM2012)\footnote{\url{www.bipm.org}} and we make use of the Jet Propulsion Laboratory (JPL) DE421 planetary ephemeris\footnote{\url{http://tmo.jpl.nasa.gov/progress_report/42-178/178C.pdf}}.  The pulsar models are based on those of Verbiest et al. (2009) and we correct the dispersion measure variations as described by Keith et al. (2012).

In order to produce data sets with the longest possible data spans we have, where possible, combined the PPTA data (which starts in the year 2005) with earlier Parkes observations of the same pulsars.  These earlier data sets have been described by Verbiest et al. (2008, 2009).  These extra data allow us to expand our data sets backwards to 1995.  The earlier data have poorer observing cadence and timing precision than the more recent data, but the most significant problem is that the data were only obtained in the 20\,cm observing band.  This restricts the precision with which the dispersion measure variations can be measured and corrected.   

\begin{figure}
\begin{center}
\includegraphics[width=8cm,angle=-90]{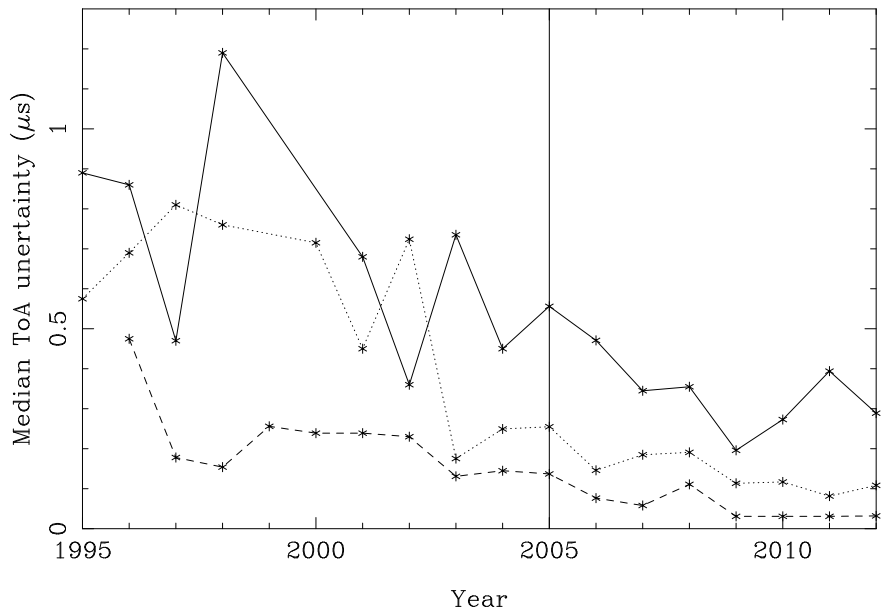}
\end{center}
\caption{Average median timing residuals per year for PSR J0437$-$4715 (dashed), J1713$+$0747 (dotted) and  PSR~J1744$-$1134 (solid)}\label{fg:median_toa}
\end{figure}

Over time our data sets have been improving as new systems (both front-end receivers and back-end instruments) have been commissioned.   To demonstrate this improvement we show, in Figure~\ref{fg:median_toa}, the median ToA uncertainty for data obtained in a given year for PSRs J0437$-$4715, J1713$+$0743 and J1744$-$1134.  Prior to the solid, vertical line the observations were from the earlier Parkes observing program.  After the vertical line the data were obtained for the PPTA project.  The timing precision continued to improve until around 2009 at which time we commissioned our current suite of instrumentation.

\begin{figure}
\begin{center}
\includegraphics[width=8cm]{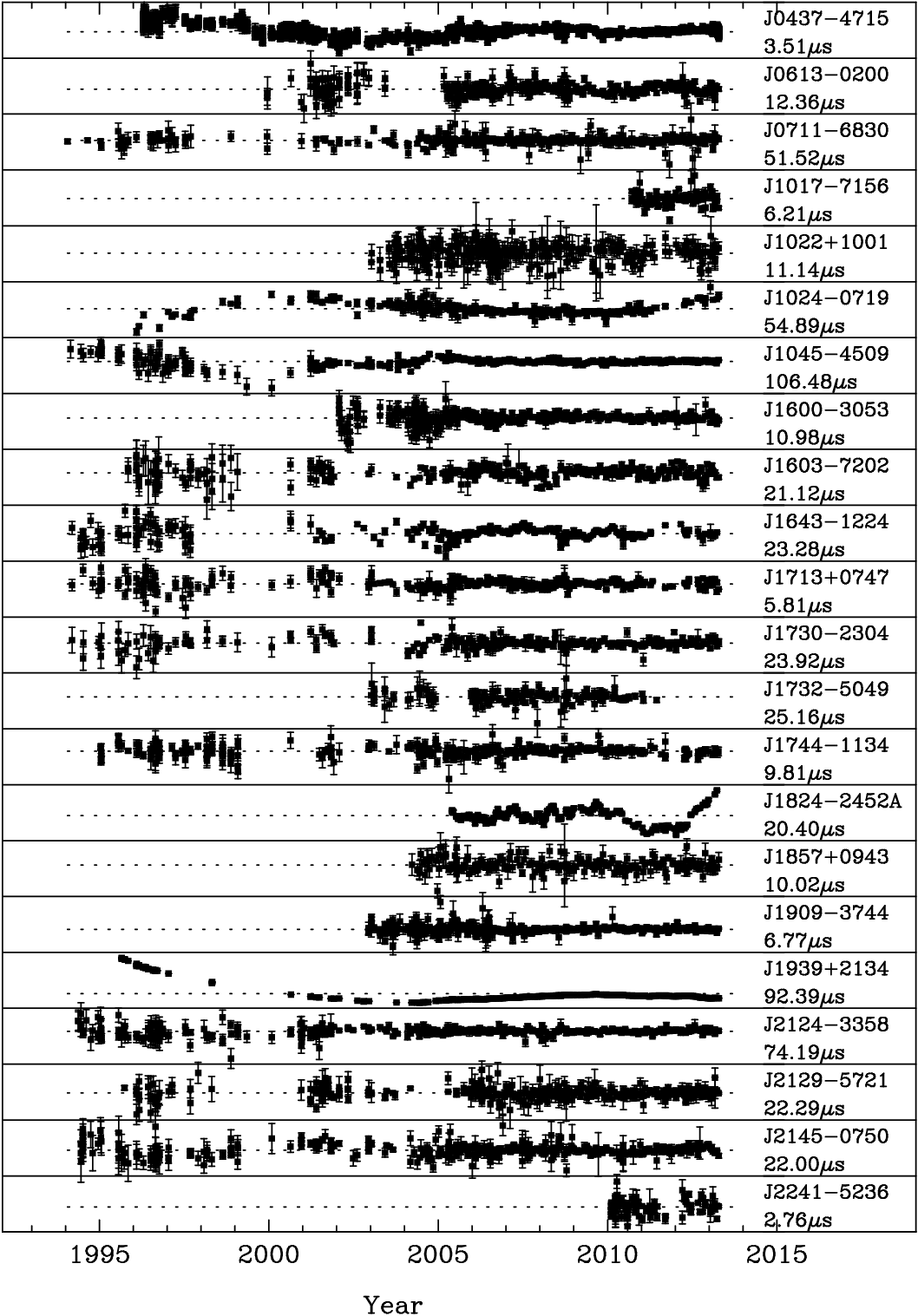}
\end{center}
\caption{Extended 20\,cm PPTA data sets.  Each panel, representing each pulsar in the PPTA sample, is scaled independently.  The value listed under the pulsar's name indicates the residual range (i.e., the highest residual minus the lowest).}\label{fg:longest}
\end{figure}

Our extended data sets are described in Table~\ref{tb:extended} and the timing residuals in the 20\,cm observing band are shown in Figure~\ref{fg:longest}.  The table lists the pulsar name, data span, total number of observations and, to give an indication of the variation in the data set, the weighted rms timing residual measured in the 20\,cm observing band.  We note that the weighted rms timing residual is often a misleading statistic.   Some pulsars scintillate strongly and so the weighted rms residual can be dominated by a few measurements (that also dominate the least-squares-fitting procedure). The rms residual is also affected by red noise in the data set and therefore depends upon data span.  Finally, the rms residual is affected by arbitrary phase jumps that often needed to be included in the timing model fit to account for time offsets between different instruments.   Details on determining these jumps are given in Manchester et al. (2013).

\begin{table}
\begin{center}
\caption{The extended Parkes Pulsar Timing Array data sets. $\sigma_{\rm 20cm}$ is the weighted rms timing residual in the 20\,cm observing band.}\label{tb:extended}
\begin{tabular}{llllll}
\hline
PSR J & First obs. & Last obs. & T$_{\rm span}$ & N$_{\rm obs}$ & $\sigma_{\rm 20cm}$\\
            &  (MJD) & (MJD)                     & (yr)                      &              & ($\mu$s) \\
\hline
J0437$-$4715 & 50191 & 56397 & 17.0 & 5160 & 0.223 \\
J0613$-$0200 & 51527 & 56395 & 13.3 & 799 & 1.177 \\
J0711$-$6830 & 49374 & 56396 & 19.2 & 729 & 1.307 \\
J1017$-$7156 & 55456 & 56395 & 2.6 & 282 & 0.684 \\
J1022$+$1001 & 52650 & 56395 & 10.3 & 755 & 1.996 \\ \\
J1024$-$0719 & 50118 & 56395 & 17.2 & 626 & 4.622 \\
J1045$-$4509 & 49406 & 56396 & 19.1 & 714 & 4.150 \\
J1600$-$3053 & 52302 & 56396 & 11.2 & 754  & 0.797 \\
J1603$-$7202 & 50026 & 56396 & 17.4 & 590 & 2.046 \\
J1643$-$1224 & 49422 & 56396 & 19.1 & 581 & 2.146 \\ \\
J1713$+$0747 & 49422 & 56396 & 19.1 & 683 & 0.399 \\
J1730$-$2304 & 49422 & 56396 & 19.1 & 514 & 1.905 \\
J1732$-$5049 & 52647 & 55725 & 8.4 & 226 & 2.246 \\
J1744$-$1134 & 49729 & 56395 & 18.2 & 637 & 0.575 \\
J1824$-$2452A & 53519 & 56395 & 7.9 & 401 & 3.661 \\ \\
J1857$+$0943 & 53087 & 56395 & 9.1 & 416  & 0.876 \\
J1909$-$3744 & 52618 & 56396 & 10.3 & 1189 & 0.186 \\
J1939$+$2134 & 49957 & 56395 & 17.6 & 489 & 4.347 \\
J2124$-$3358 & 49490 & 56395 & 18.9 & 713 & 2.677 \\
J2129$-$5721 & 49987 & 56396 & 17.5 & 576 & 1.110 \\ \\
J2145$-$0750 & 49518 & 56396 & 18.8 & 751& 1.066 \\
J2241$-$5236 & 55235 & 56396 & 3.2 & 292 & 0.320 \\
\hline
\end{tabular}
\end{center}
\end{table}

\section{Current research}


Members of the PPTA team are working on all aspects of the project - from improving the instrumentation at Parkes to developing the algorithms required for us to achieve the main aims of the project.  Here we first describe improvements being made in the observing system.  We then highlight current research related to the three main project goals.

\subsection{Improving the PPTA data sets}


\begin{figure}
\begin{minipage}{8cm}
\includegraphics[width=5cm,angle=-90]{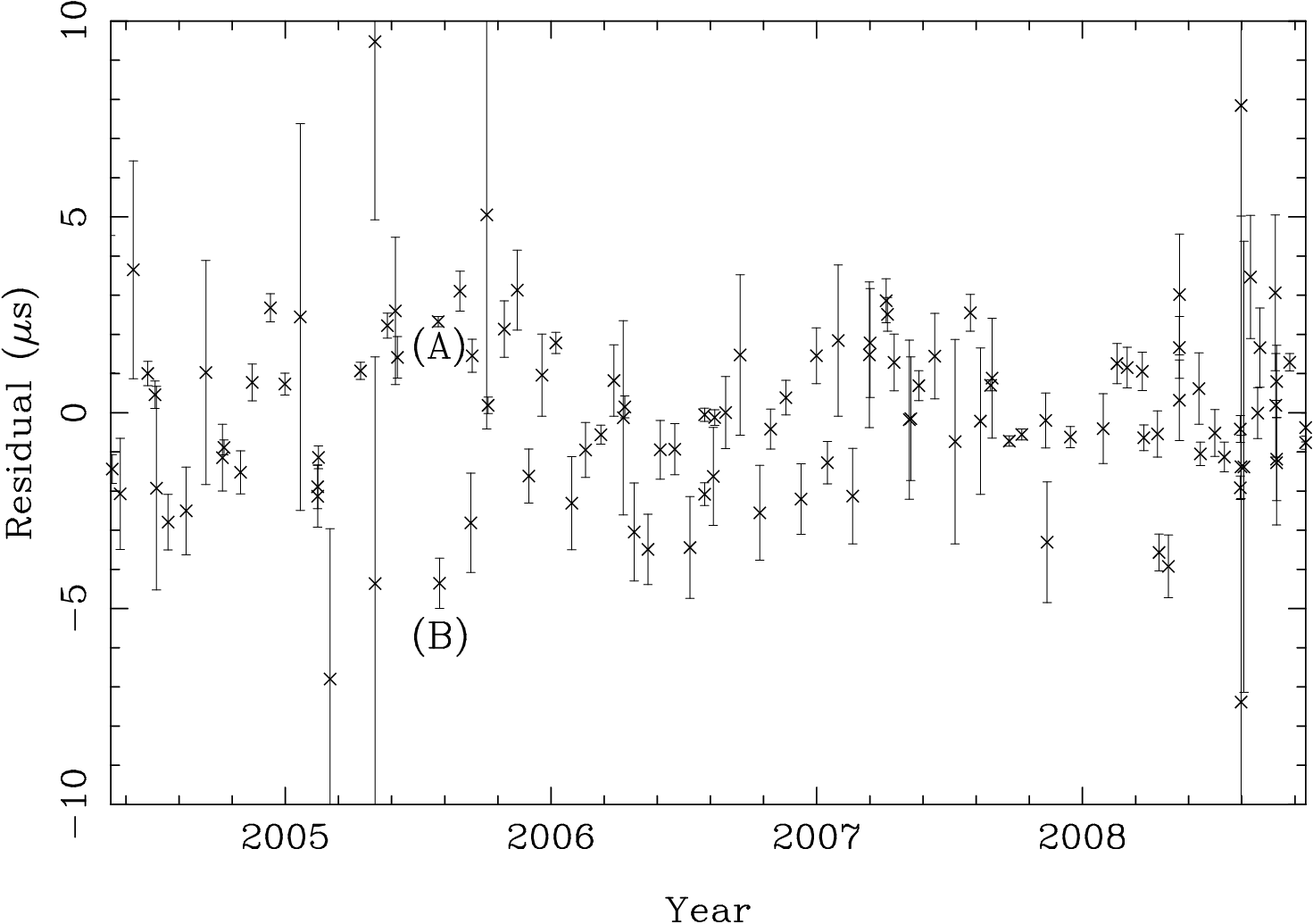} 
\end{minipage} 
\begin{minipage}{8cm}
\includegraphics[width=5cm,angle=-90]{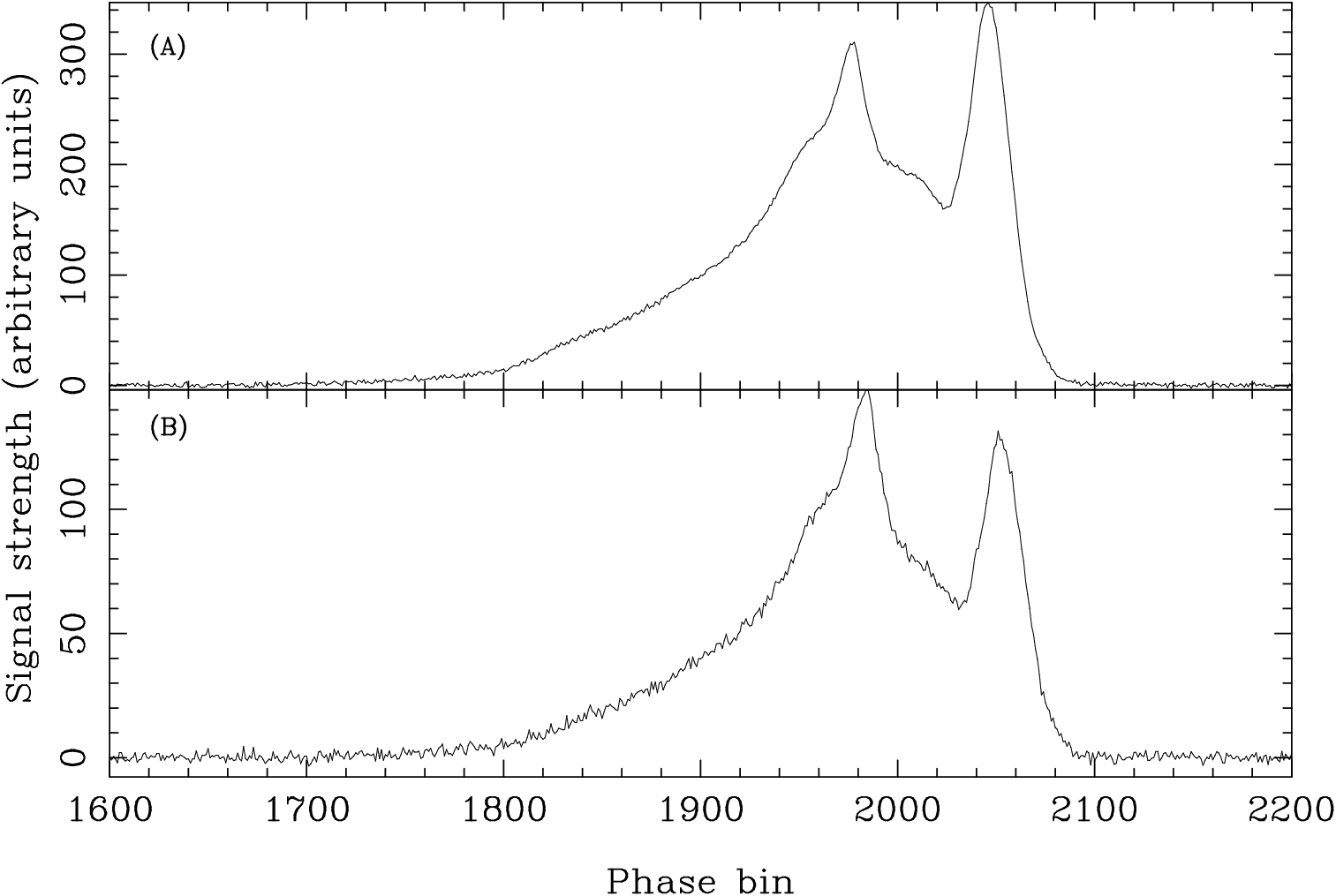} 
\end{minipage}
\caption{(left panel) Timing residuals for PSR~J1022$+$1001 in the 20\,cm observing band.  Two observations separated by two days are indicated by the labels (A) and (B). (right panel) the profiles for the observations indicated.}\label{fg:1022}
\end{figure}

\begin{figure}
\begin{minipage}{7cm}
\includegraphics[width=7cm]{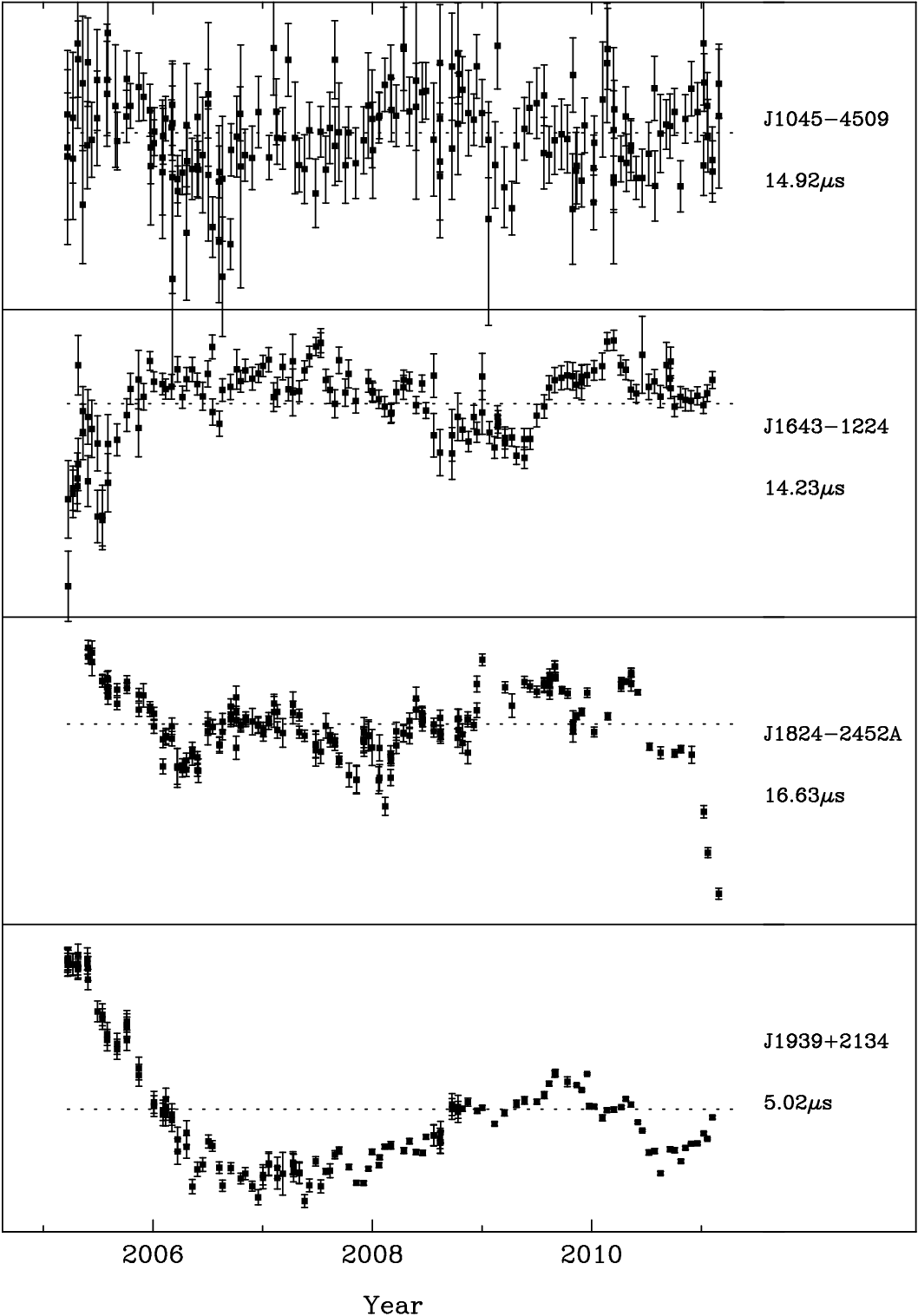} 
\end{minipage}
\begin{minipage}{7cm}
\includegraphics[width=6cm,angle=-90]{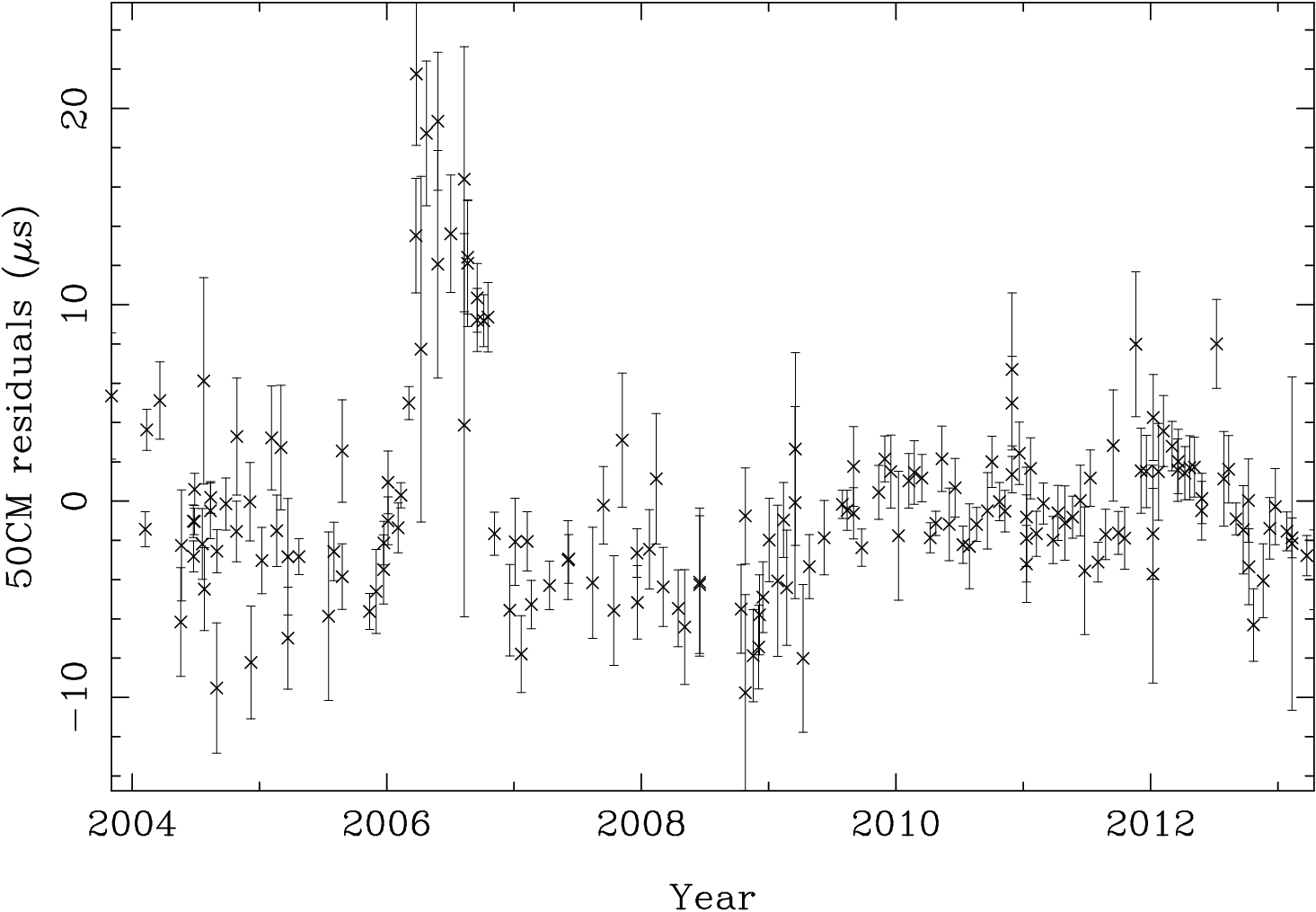} 
\end{minipage}
\caption{(left panel) A sample of 20\,cm data sets that have been corrected for dispersion measure variations, but are still significantly affected by low frequency noise. (right panel) The timing residuals in the 50\,cm band for PSR~J1603$-$7202.}\label{fg:redNoise}
\end{figure} 

The initial intention for the PPTA project was to obtain timing residuals for around 20 pulsars with rms timing residuals of $\sim$100\,ns.  This has not been achieved.  Many of the pulsars are relatively weak and with the available observation times and receiver systems we are only able to achieve timing precisions between $\sim$500\,ns and $\sim 1\mu s$.  However, some of our data sets are affected by unexplained white- and red-noise processes. Removing the cause of such excess noise would significantly improve these data sets.  In this section we describe the current status of current research to study such excess noise. 

Oslowski et al. (2011) analysed 25 hours of observations of PSR J0437$-$4715 and showed that the timing residuals had a standard deviation around four times the expected value given the arrival time uncertainties.  This was explained as being caused by the intrinsic variability of the pulse shape and they showed that, in the 20\,cm band, timing precision better than 30--40\,ns in a 1 hour observation is highly unlikely, regardless of future improvements in telescope sensitivity. 
 
In Table~\ref{tb:pulsars}, PSR~J1022$+$1001 has a median error bar size in the 20\,cm band of 0.8\,$\mu$s.  The timing residuals show no excess red-noise and yet the weighted rms timing residual is 1.2\,$\mu$s and the unweighted rms is 2.4\,$\mu$s (Figure~\ref{fg:1022}).  In the figure two observations are labelled as (A) and (B).  They are separated in time by only two days, but the residuals are $\sim$\,7$\mu$s apart.  The reason for this is shown in the right hand panel of Figure~\ref{fg:1022}.  The profiles in the two observations differ.  In (A) the leading pulse component is weaker than the trailing component.  In (B) the reverse is observed. The reason for this variability is not yet understood, but is likely to be caused by calibration errors (see, van Straten et al. 2013), intrinsic pulse shape variability (Kramer et al. 1999) and/or scintillation and pulse-shape evolution across the band (Ramachandran \& Kramer 2003).  

Many of our pulsar data sets indicate the presence of an unmodelled red noise process.  Manchester et al. (2013) reports that approximately half of the pulsar sample have a significant $\ddot{\nu}$ measurement.  The most extreme cases are shown in Figure~\ref{fg:redNoise} where we show the timing residuals after correction for dispersion measure variations.   If this red noise has similar properties to the timing noise studied by Lyne et al. (2010) then it may be possible to correct this timing noise by searching for correlations with pulse shape variations.  To date, we have not been successful in our attempts to do this.
 
The timing residuals for PSR~J1603$-$7202 show another unmodelled process (right panel in Figure~\ref{fg:redNoise}).  The excess that lasts for around 250\,d during the year 2006 has been explained by Keith et al. (2013) as a discrete change in the dispersion measure of  $\sim 2 \times 10^{-3}$\,cm$^{-3}$pc possibly caused by an extreme scattering event. 

Improving our calibration procedures and attempting to correct for red timing noise will significantly improve our current data sets.  However, it is also necessary to continue to improve our hardware systems.  We are currently proposing a wide-band receiver system that will cover the band from 0.7 to 4\,GHz. The entire band will be directly digitised in the focus cabin and processed by a GPU-based processor.  It is hoped that such a system will be commissioned within two years. 

\subsection{Gravitational wave detection}

\begin{figure}
\begin{minipage}{7cm}
\includegraphics[width=6cm,angle=0]{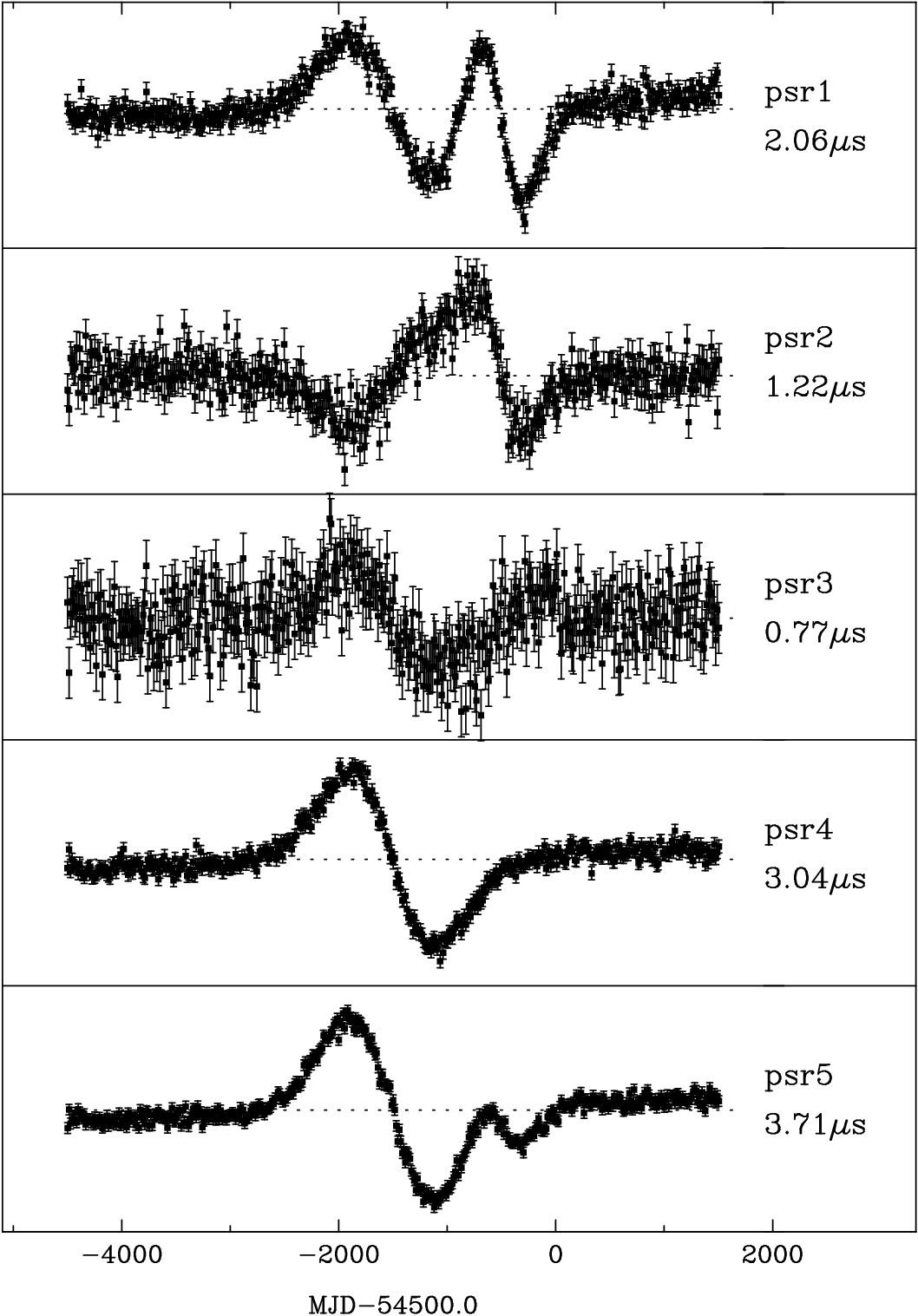} 
\end{minipage}
\begin{minipage}{7cm}
\includegraphics[width=6cm,angle=-90]{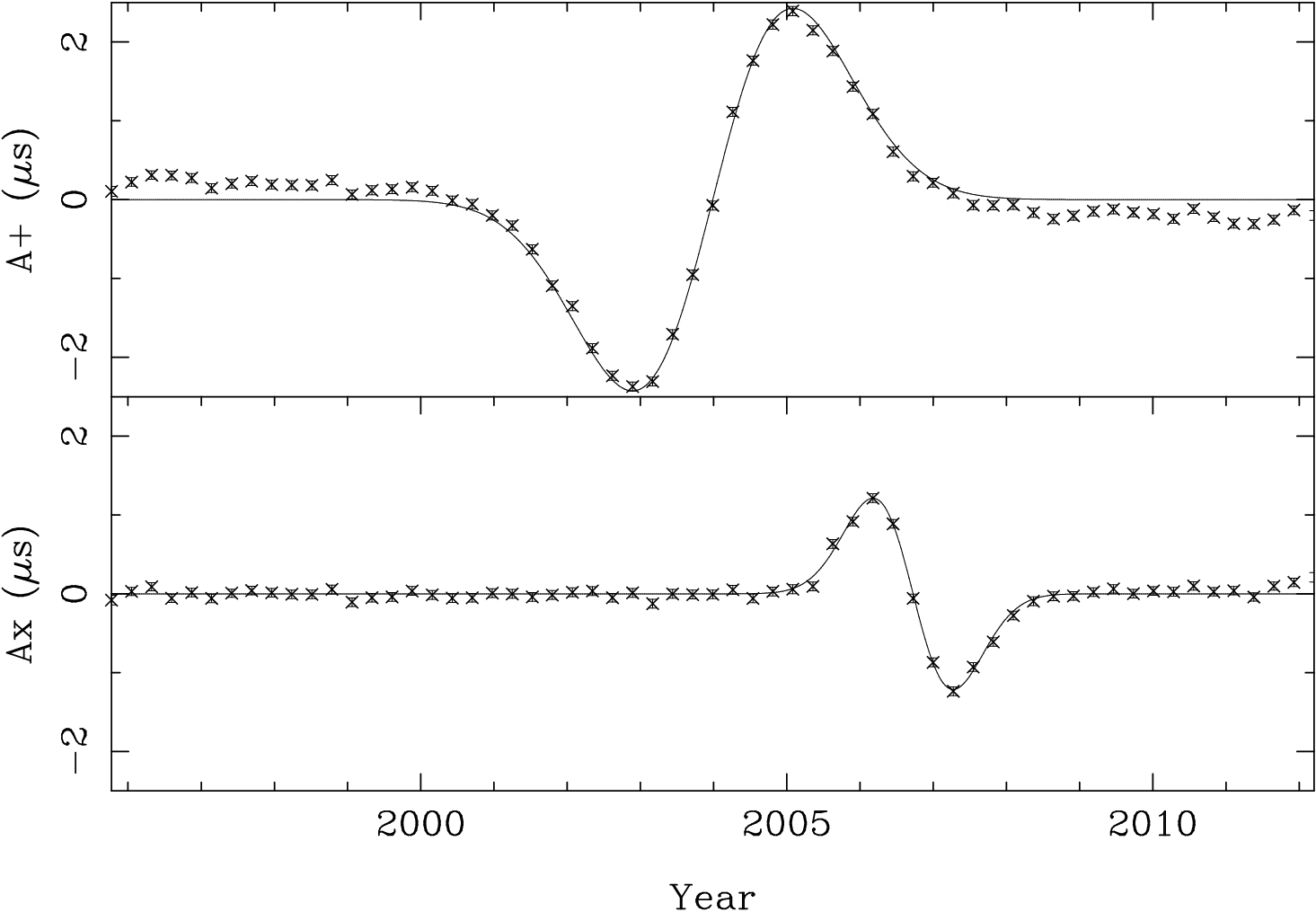} 
\end{minipage}
\caption{The left-hand panel shows the simulated timing residuals for five pulsars that are affected by a gravitational wave burst.  The functional form of the burst in the two polarisation states A$_+$ and A$_\times$ is shown as the solid lines in the right hand panel. The \textsc{tempo2} fit for  A$_+(t)$ and A$_\times(t)$ is shown in the right-hand panel with error bars. }\label{fg:burst}
\end{figure}

New theoretical calculations are suggesting that the detectable gravitational wave signal is unlikely to be an isotropic, stochastic gravitational wave background (see, for instance, Ravi et al. 2012).  We therefore require gravitational-wave detection algorithms that are sensitive to backgrounds, individual continuous wave sources, evolving sources, burst events or memory events.   We have recently attempted to simplify the search for these various types of wave, by noting that the pulsars in the PPTA act as individual elements of a giant gravitational wave telescope.  By suitably weighting the timing residuals from each pulsar we can ``point" this gravitational wave telescope to a particular direction in the sky and hence obtain the functional form of the gravitational wave signal from that direction.  We have recently updated the \textsc{tempo2} software package to produce time series for the two gravitational wave polarisation states ($A_+(t)$ and $A_\times(t)$) from a specified sky direction.  Separate algorithms can then be applied to search the time series corresponding to different sky positions for the signatures of burst events, individual sources or memory events.

As an example we show, in the left-hand panel of Figure~\ref{fg:burst}, five simulated data sets that contain a unrealistic, but instructional, gravitational wave burst event.  In the right-hand panel of Figure~\ref{fg:burst} we show the resulting $A_+(t)$ and $A_\times(t)$ time series (data points with errors) that clearly recover the simulated burst (solid lines). The resulting fit is not perfect as it is impossible to measure the linear or quadratic component of a burst event that lasts longer than the data span.  This is because the pulsars' intrinsic pulse periods and time derivatives are unknown.   The $A_+(t)$ and $A_\times(t)$ time series are therefore constrained within the fit not to include a quadratic polynomial.

The method described above can be used to search for individual sources of gravitational waves.  However, this method is not suitable for a gravitational wave background.  We are therefore improving the algorithm described by Yardley et al. (2011).  Our algorithm is still being developed, but is based on forming cross-power spectra for each pair of pulsars using the Cholesky method for dealing with steep, red noise (Coles et al. 2011).  These cross-power spectra are used to form the covariance between two data sets.  The correlations between pairs of pulsars as a function of the angle between the pulsars are expected to follow the prediction of Hellings \& Downs (1983).  Our algorithm is still work-in-progress, but we have successfully applied our method to the International Pulsar Timing Array data challenge\footnote{\url{http://www.ipta4gw.org/?page_id=214}}.  

\subsection{Improving the solar system ephemeris}

\begin{figure}
\begin{center}
\begin{minipage}{7cm}
\includegraphics[width=6cm,angle=0]{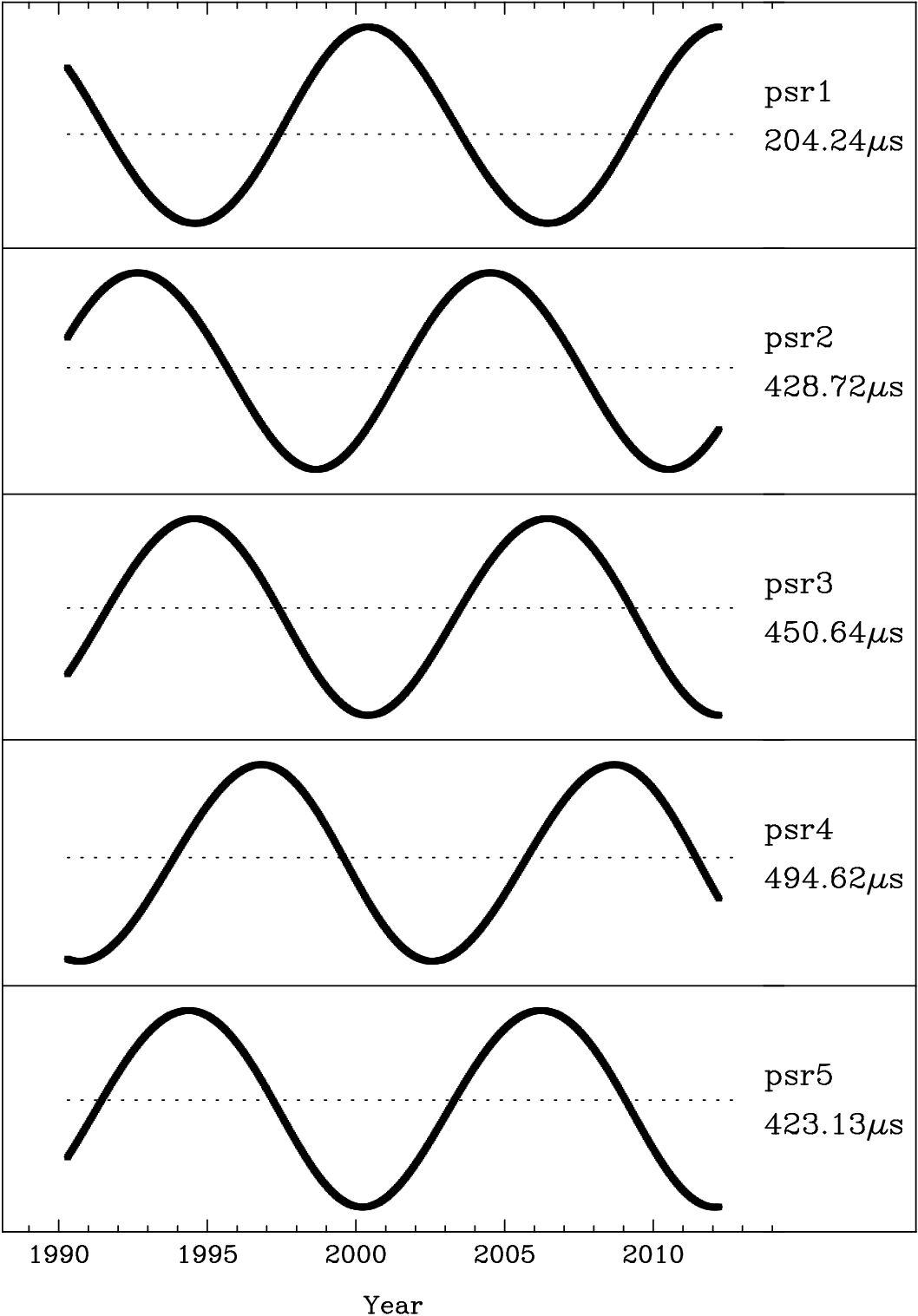} 
\end{minipage}
\begin{minipage}{7cm}
\includegraphics[width=6cm,angle=-90]{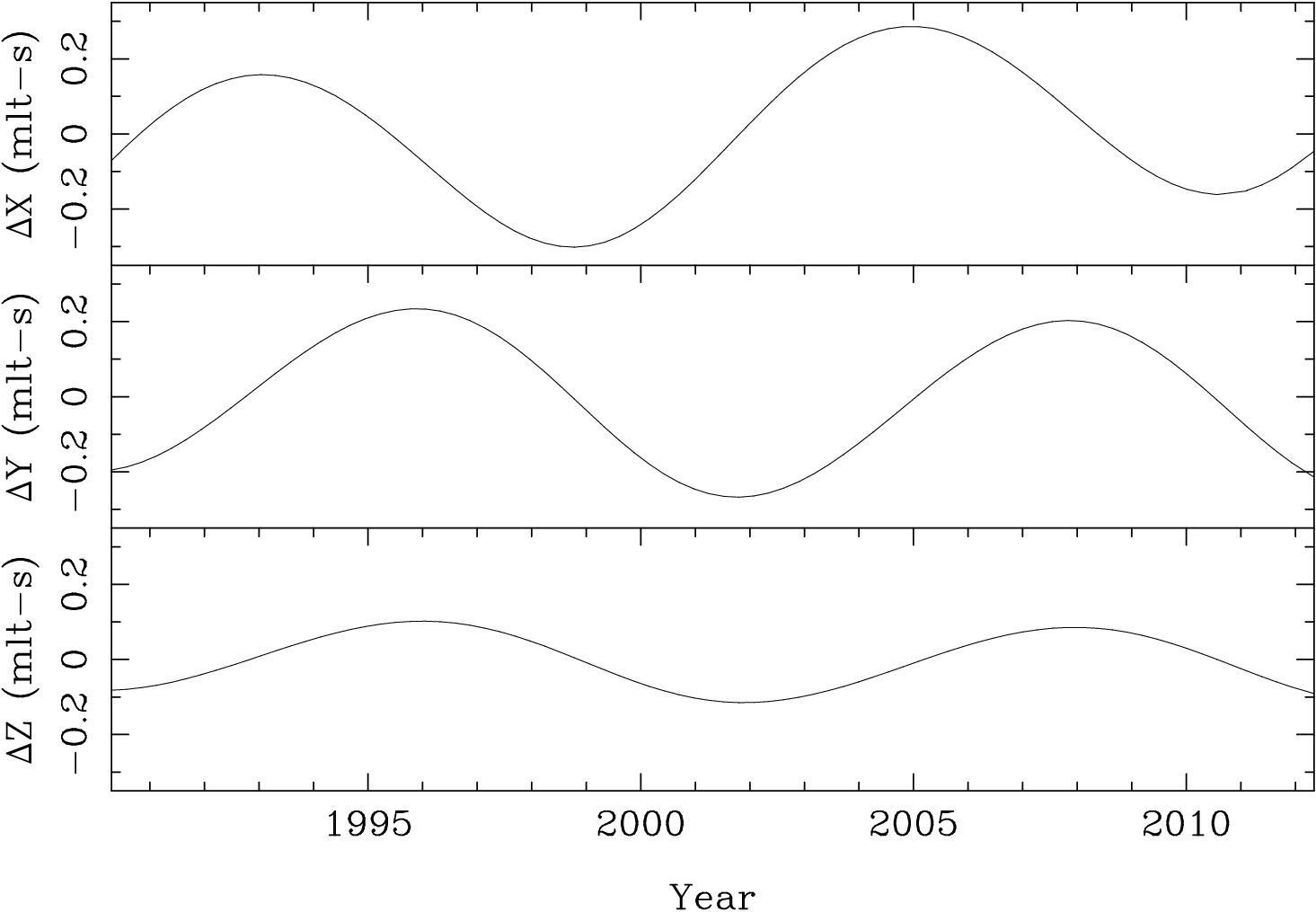} 
\end{minipage}
\end{center}
\caption{Left-hand panel shows the induced timing residuals due to an incorrect mass of Jupiter of $\Delta M_J = 10^{-7}M_\odot$.  The right-hand panel displays the offset in the observatory-SSB position in X, Y and Z as a function of time.}\label{fg:dxdydz}
\end{figure}

Champion et al. (2010) searched for the signatures of incorrect mass estimates of the planetary systems in our solar system.  This method relies on knowledge of the relative positions as a function of time of the planetary system, the Earth and the solar system barycentre (SSB). This method is therefore not applicable to searching for unknown masses in the solar system.  We have therefore updated the \textsc{tempo2} software to measure offsets in the estimate of the Earth's position with respect to the SSB using a PTA data set.  The initial estimate of the Earth's position with respect to the SSB is obtained from a planetary ephemeris.  Any error, or omission, in that ephemeris will therefore lead to an incorrect estimate of the Earth-SSB vector.  Inspection of time series of the error in the three components of the Earth-SSB vector allows unknown masses to be identified.  As an example we simulate data sets that include an error in the mass of the Jovian system of $\Delta M_J = 10^{-7}M_\odot$.  This mass error could easily be measured using the Champion et al. (2010) approach, but here we make no assumption about the nature of the error in the planetary ephemeris.  We assume that five pulsars have been observed since 1990.  The output of the \textsc{tempo2} global fit provides the offset of the Earth-SSB vector from the planetary ephemeris prediction (Figure~\ref{fg:dxdydz}).  These offsets could subsequently be searched to identify the orbital period of the unknown mass and the orbital angle with respect to the ecliptic plane,  thereby determining the position and orbit of the unknown objects.  

This procedure can be generalised and used to determine the position of any telescope at any position in the solar system.  We have recently used PPTA data to show how this method can be used to navigate a space craft through the solar system (Deng et al. submitted).

\subsection{Pulsar-based time standards}

In Hobbs et al. (2012b) we describe new updates to \textsc{tempo2} that allow the signal common to all pulsars to be identified given a PTA data set.  This method was applied to the PPTA data sets and we recovered the known offsets between the world's best time standards: International Atomic Time and Terrestrial Time as realised by the Bureau International des Poids et Mesures  (BIPM).   One major issue with our method is that we assume that the noise in the data sets has the same statistical form throughout the observations.  However, as described in Manchester et al. (2013), we are unable to correct our earliest data sets for dispersion measure variations.  The timing residuals for most pulsars are therefore affected, for the earliest observations, by dispersion measure variations and, for the most recent observations, by other noise processes such as intrinsic pulsar timing noise.  We are currently enhancing the Coles et al. (2011) Cholesky routines to account for non-stationary noise processes and will use the new routines to improve our pulsar-based time standard.

\section{Outreach}

Some of the PPTA observations are carried out by high school students as part of the PULSE@Parkes education program (Hobbs et al. 2009).  This program provides high school students with two hours of observing per month.  So far more than 1000 students from Australia, Japan, the USA, the Netherlands, England and Wales have been part of the program.

\section{Conclusion}

The Parkes Pulsar Timing Array is continuing to obtain high quality pulsar timing observations of $\sim$ 20 millisecond pulsars.  These data sets are leading to exciting new research on topics as diverse as studying the solar wind to searching for bursts of gravitational waves to navigating space craft through the solar system. The data sets are also being combined with observations from Europe and North America to form the International Pulsar Timing Array data sets.   With further development to the hardware systems at Parkes it is likely that these observations will continue to contribute to timing array data sets into the Square Kilometre Array era.

\end{document}